\documentclass[aps,a4paper,superscriptaddress,twocolumn]{revtex4}
\usepackage{tensor}
\usepackage{graphicx}
\graphicspath{ {images/} }
\usepackage{amsmath}
\usepackage{amssymb}
\usepackage{enumerate}
\usepackage{subfigure}
\usepackage{tabularx}
\usepackage{lipsum} 
\usepackage{float} 
\usepackage[colorlinks=true, pdfstartview=FitV, linkcolor=blue, citecolor=red, urlcolor=black, breaklinks=true]{hyperref}
\usepackage{xcolor}
\newcommand{\be}{\begin{equation}}
\newcommand{\ee}{\end{equation}}
\newcommand{\ben}{\begin{eqnarray}}
\newcommand{\een}{\end{eqnarray}}
\newcommand{\bes}{\begin{subequations}}
	\newcommand{\ees}{\end{subequations}}
\def\bal#1\eal{\begin{align}#1\end{align}}

\newcommand{\sech}{{\rm sech}}
\newcommand{\LL}{{\mathcal L}}

\newcommand{\pu}{\mathrm{\partial_{\mu}}}

\newcommand{\Pu}{\mathrm{\partial^{\mu}}}

\newcommand{\pb}[1]{\ensuremath{\partial_{#1}}}

\begin{document}
	\title{Impurity-doped scalar fields in arbitrary dimensions}
	
	\author{D. Bazeia}\affiliation{Departamento de F\'\i sica, Universidade Federal da Para\'\i ba, 58051-970 Jo\~ao Pessoa, PB, Brazil}
	\author{M. A. Liao}\affiliation{Department of Physics and Astronomy, University of Pennsylvania, Philadelphia, PA 19104, USA}\affiliation{Departamento de F\'\i sica, Universidade Federal da Para\'\i ba, 58051-970 Jo\~ao Pessoa, PB, Brazil}	\author{M. A. Marques}\affiliation{Departamento de Biotecnologia, Universidade Federal da Para\'iba, 58051-900 Jo\~ao Pessoa, PB, Brazil}
	
	\begin{abstract}
		We investigate the presence of localized structures for relativistic scalar fields coupled to impurities in arbitrary spatial dimensions. Such systems present spatial inhomogeneity, realized through the inclusion of explicit coordinate dependence in the Lagrangian. It is shown that, in stark contrast to the impurity-free scenario, Derrick's argument does not present a strong hindrance to the existence of stable solutions in this case. Bogomol'nyi equations giving rise to global minima of the energy are found, and some of the ensuing BPS configurations are presented.
			\end{abstract}
	
	
	\maketitle

Several works in field theory and condensed matter physics have explored the effect of spatial inhomogeneities; see, for example, ~\cite{Anderson, Benincasa, Ash, Chou, Liu, Goryo, RMP,Li, Hook, Fan, Neupert} and references therein. These inhomogeneities may be represented by the addition of coordinate-dependent terms to the Lagrangian density, thus incurring changes to the equations of motion, representing interactions between the fields and impurities. This generalization allows for more realistic treatment of physical situations in which uniformity of spacetime may not be a reasonable assumption. Systems
 in which impurities have been investigated also include superconductors, where they have been coupled to topological  vortices~\cite{Tong, Ashcroft,Cockburn, Yin,Existence, BLM}, Bose-Einstein condensates~\cite{Mancini, Catani, Hu, Akram}, holography~\cite{Kachru,BenincasaII, Harrison, Seo, Latychevskaia, Evans}, Fermi liquids~\cite{Evans, Jensen} and many others.
 
 The coupling of a scalar field theory with impurities in $1+1$ Minkowski spacetime has been investigated in Refs.~\cite{Kivshar,Malomed, Fei, AdamI, Manton,AdamII}, and many interesting new properties have been found. However, every work thus far conducted in the subject seems to have been limited to two spacetime dimensions, unless one takes the target space more complicated than the one to be considered in the present work; see, e.g., Refs. {\cite{R1,R2}}. In the impurity-free or {\it homogeneous} case, this dimensional limitation is natural in light of the well-known Derrick's Theorem~\cite{Derrick}, which forbids the existence of stable static solutions in real scalar field theories governed by a  {\it standard} Lagrangian density $\LL_{st}=(1/2)\pu\phi\Pu\phi-U(\phi)$. However, one should not assume that Derrick's theorem is automatically valid in the presence of impurities, since such a generalization changes the variation of the energy functional with respect to a rescale transformation. In this work, we show that not only are stable solutions possible when a standard theory is coupled to a impurity, but the coupling may be chosen in a way that allows for the presence of BPS solutions in the system. These solutions, named after Bogomol'nyi~\cite{bogo}, Prasad and Sommerfield~\cite{ps}, are global minima of the energy functional for given boundary conditions. In the homogeneous case, energy minimizers have been found for the restriction of the theory under some symmetry for non-standard Lagrangians~\cite{A1,stable, effective} and for standard theories on curved backgrounds~\cite{curved,stableDwall}. However, solutions which satisfy the BPS property without the assumption of some symmetry have, to the best of our knowledge, never been found before in scalar field theories in more than two spacetime dimensions. This difference suggests that solutions with a lower energy but lacking symmetry do not exist.

The inclusion of impurities (see, e.g., Refs. \cite{Tong,Ashcroft}) is somewhat similar to the addition of spin-orbit couplings, which includes spatial dependence and first-order derivative, being tantamount to the Dzyaloshinskii-Moriya (DM) interaction \cite{Dz,Mo}. This opens several other possibilities of investigations, in particular, in the case of N\'eel and Bloch domain walls and skyrmions in magnetic materials in the presence of the DM interaction \cite{DM1,DM2}, the manipulation of the magnetic order \cite{soc} and the study of Bose-Einstein (BE) condensates with the inclusion of spin-orbit coupling \cite{SOC1,SOC2}. It may also appear as a response of the demagnetization field and the interlayer DM interaction, which contributes to break the Bloch wall chirality related to the film thickness in magnetic multilayers \cite{inter}. In BE condensates, a route of practical interest may be connected to the study developed in \cite{ABC}, in which the use of similarity transformation was considered to connect nonlinear Schr\"odinger equation with equation of motion for scalar field.
The intrinsic interest related to relativistic fields then adds to the possibility of applications to non-relativistic systems such as magnetic materials and condensates to compose the basic motivation of the present study.

The Derrick's Theorem~\cite{Derrick} is an important obstruction to the study of relativistic scalar fields in arbitrary dimensions. The argument is built upon the variation of the static energy functional given the rescale $\textbf{x}\to \lambda \textbf{x}$ $(\lambda\in\mathbb{R})$. In a standard $(D+1)$-dimensional theory, this procedure implies $DE_p-(2-D)E_g=0$, where $E_g$ and $E_p$ are given, respectively, by integration of $(\nabla\phi)^2/2$ and $U(\phi)$. For $D\geq 3$, this condition leads to $\delta^2 E <0$, thus implying instability, while the $D=2$ case can only evade the theorem if $E_p=0$. If impurities are added to the system, new terms must be added to this calculation. We can think of a scalar field coupled to  impurities through a sum of terms of the form $\LL^{(k)}=-c_{k}^a(\phi,\mathbf{x})(\pb{a}\phi)^k(\mathbf{x})$. Together, these terms give a contribution to $dE_\lambda/d\lambda|_{\lambda=1}$ as
 \begin{equation}\label{condition}
 \begin{split}
&\sum_{k}\int d^Dx\left\{\left[(\pb{a}\phi)^k(\nabla c_{k}^a(\phi,\mathbf{x}))\cdot\mathbf{x}\right]\right.\\
&\big. \;\;\;\;\;- (k-D)\LL^{(k)}(\phi,\pb{a}\phi,\mathbf{x})\big\}, \end{split}
 \end{equation} 
 which must equal $DE_p-(2-D)E_g$ for stability. Although the calculation above takes only first and zeroth derivative coupling, it is straightforward to verify that terms involving $k$-th derivative coupling work in a similar way. The above condition is far more permissive than the one encountered in the homogeneous case. Moreover, since the interaction between scalar field and impurity may be repulsive as well as attractive, no sign restriction is placed upon the $c_k^a$ in general, thus broadening the class of models in which $d^2E_{\lambda}/d\lambda^2\rvert_{\lambda=1}$ gives a positive result. Indeed, stable static solutions may be found even in the simplest possible case, namely the one where $c_{0}=c_0(\phi)$ is the only nonzero coefficient. This case is mathematically equivalent to that of a Lagrangian with a non-canonical potential $U(\phi,\mathbf{x})$. The $\textbf{x}$-dependence in the potential modifies the vacuum structure of the theory, giving rise to energy minimizers that are nontrivial functions of the coordinates. 
 
 Systems of the kind discussed above have been found during our investigations, but in the present Letter we shall focus on defects that possess the BPS property. \textcolor{black}{Inspired by Refs.~\cite{AdamI, AdamII}, we attempt to develop BPS equations for a single real scalar field in arbitrary dimensions. We consider} a Lagrangian of the form $\LL=\LL_{st} + {\cal L}_{im}(\mathbf{x}, \phi,\pb{a}\phi)$, to account for the presence of impurity. To find a Bogomol'nyi bound, we introduce $D$ real parameters $\alpha_k$ such that $\alpha^2_1+...+\alpha_D^2=1$. \textcolor{black}{In order to work out a Bogomol'nyi procedure~\cite{bogo}, we consider completing squares in the static energy functional of the theory to find that BPS solutions are possible. This suggests that}
 \begin{equation}
 {\cal L}_{im}=\sum\limits_{k=1}^{D}\left[\sigma_k(\mathbf{x})\pb{k}\phi-\sqrt{2U}\alpha_k\sigma_k(\mathbf{x}) - \frac{(\sigma_k)^2}{2}\right],
 \end{equation}
where each $\sigma_k(\mathbf{x})$ is an impurity function. Together, they play the role of inhomogeneity, and their interaction with the fields of the theory is dictated by ${\cal L}_{im}$. {\color{black} The potential $U=U(\phi)$ is supposed to be a non-negative function of the scalar field.} Also, we note the formal similarity between each additive term in this Lagrangian and the know results for an impurity-doped scalar theory in one spatial dimension. Indeed, in the particular case in which $\sigma_k=\sigma_k(x_k)$ for every $k$, these terms have the same form presented in~\cite{AdamI} for the one-dimensional case. Moreover, in the case of localized impurity, ${\cal L}_{im}$ vanishes asymptotically, thus recovering the canonical theory at a distance large compared to the range of the inhomogeneity. 

\textcolor{black}{In the above model, the time-dependent equation of motion can be written in the form
\begin{equation}\label{eom}
    \pu\Pu\phi + \pb{k}\sigma_k + \left(1 + \frac{\alpha_k\sigma_k}{\sqrt{2U}}\right)U_\phi=0,
\end{equation}
where $U_\phi = dU/d\phi$. For static configurations, it becomes
\begin{equation}\label{eomstatic}
    \pb{k}\pb{k}\phi = \pb{k}\sigma_k + \left(1 + \frac{\alpha_k\sigma_k}{\sqrt{2U}}\right)U_\phi.
\end{equation}
}\textcolor{black}{Moreover, the accompanying energy is}
 \begin{align}\label{Energy}
E=&\frac{1}{2}\sum\limits_{k=1}^{D}\int d^Dx\left(\pb{k}\phi-\sigma_k-\alpha_k\sqrt{2U}\right)^2  \nonumber\\
&+ \int d^Dx\sqrt{2U}\left(\sum\limits_{k=1}^{D}\alpha_k\pb{k}\phi\right).
\end{align}
\textcolor{black}{The potential $U(\phi)$ is supposed to engender spontaneous symmetry breaking, as is often the case in theories of topological defects.} We may write $U=W_{\phi}^2$, where $W_\phi=dW/d\phi$ and  $W=W(\phi)$ is an auxiliary function defined in analogy to a superpotential. If the last integral in~\eqref{Energy} is completely specified by the boundary values of the problem, its value determines a Bogomol'nyi bound relative to those boundary conditions. An energy minimizer must solve the first-order equations 
 \begin{equation}\label{bps}
\pb{k}\phi=\sigma_k+\alpha_k\sqrt{2}\ W_{\phi},\;\;\;\;\;k=1,2,...,D,
 \end{equation} 
 and its energy is determined by the last integral in~\eqref{Energy}, whose integrand is \textit{formally} equivalent to a $D$-dimensional divergence. Indeed, it is
 \be\label{TS}
E_s=\sqrt{2}\int d^D x\;\sum\limits_{k=1}^{D}\alpha_k \partial_k W.
 \ee In one spatial dimension, these results agree with the investigations of Refs.~\cite{AdamI,AdamII}, as expected.

A mathematically simple, yet physically important class of BPS solutions is found by setting $U\approx 0$, which may be seem as limiting case valid for a wide class of models. This approximation gives rise to the system of equations $\pb{k}\phi=\sigma_k$, in which the solutions are determined solely by the $\sigma_k$. Localized solutions of these equations are only possible if the impurities fall to zero at spatial infinity, so that the conditions $\pb{j}\phi\to 0$ may be compatible with the first-order equations. In order for this approximation to be valid, the potential energy $E_p$  must be negligible for the problem at hand. This argument is in fact similar the one that lead to one of the first BPS solutions ever found: the Prasad-Sommerfield monopole~\cite{ps}. In that case, the Bogomol'nyi limit arises for $\mu\to 0$, where $\mu$ is a parameter of the potential $V(\phi)=\frac{\mu^2}{2}(1-\phi^2)^2$; physically, this limit corresponds to a Higgs-particle with vanishing mass~\cite{bogo}. That reasoning is also valid here for the same potential, with the same physical interpretation. The Bogomol'nyi bound corresponding to this case is $E\geq 0$. In the homogeneous scenario this value would only be possible for a trivial (i.e., constant) vacuum solution, but the introduction of inhomogeneities in our system gives rise to a non-positive definite term in the energy density $\rho=-\LL$, thus opening up the possibility of a nontrivial energy minimizer. This is one way in which the addition of impurities is observed to modify the process of energy minimization. The procedure has not been previously reported in the literature, and it seems to open distinct lines of applications. 
 
\begin{figure}[h]
 	\centering
 	\includegraphics[width=4.2cm, height=4cm, clip]{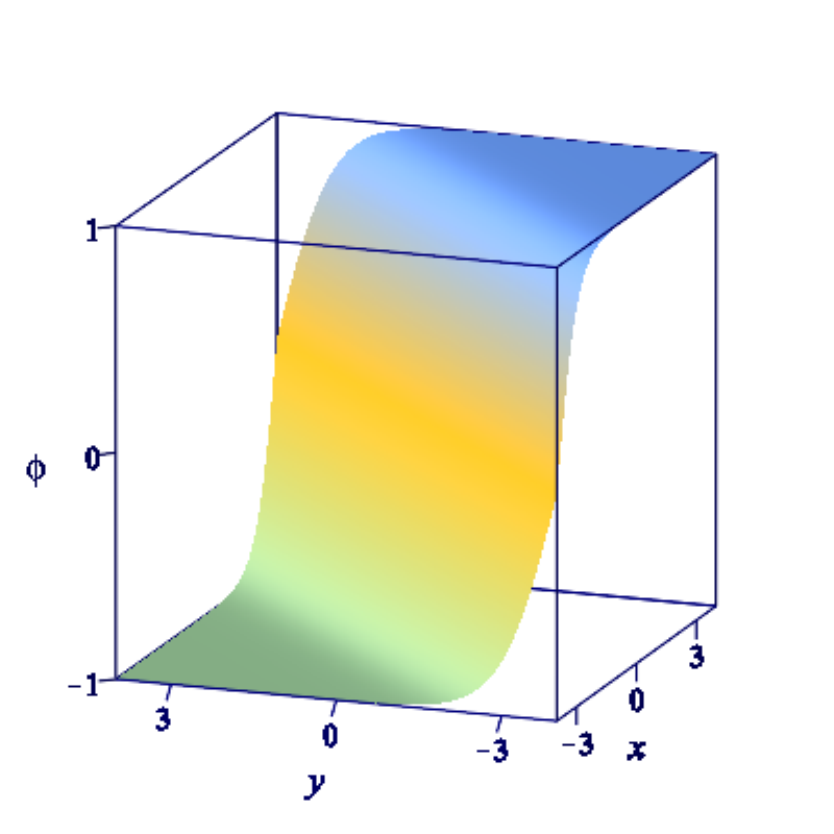}
 	\includegraphics[width=4.2cm, height=4cm, clip]{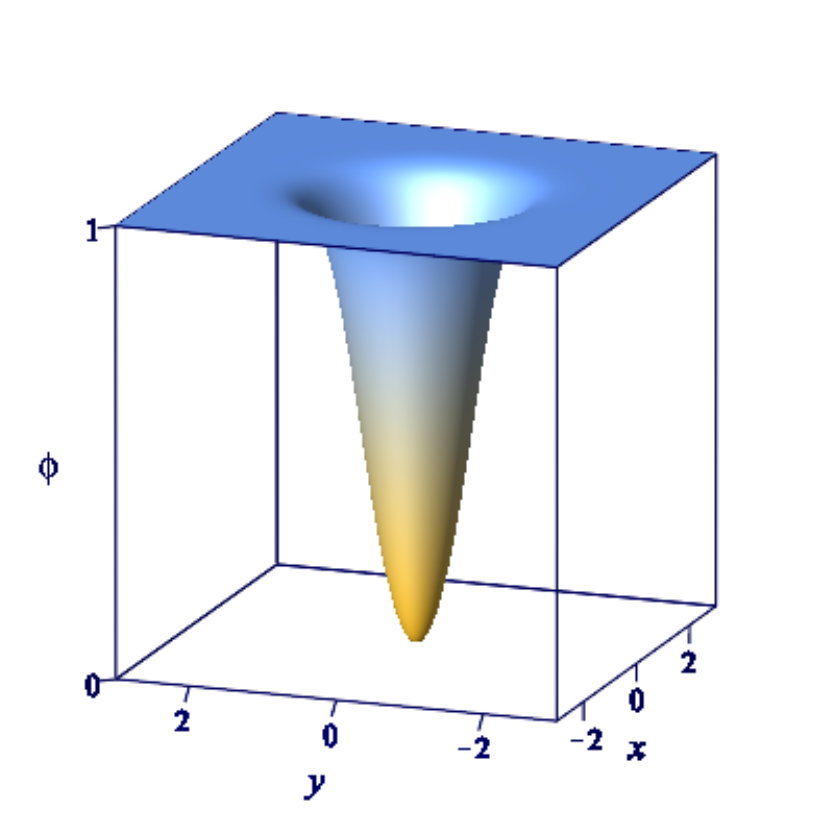}
 	\caption{Solutions $\phi=\tanh(x-y)$ (left) and  $\phi=\tanh(x^2+ y^2)$ (right) of Eqs.~\eqref{eq3}, with distinct impurities.}
 	\label{fig1}
 \end{figure}

 Let us then present some concrete examples. In three spacetime  dimensions, the model has two parameters, $\alpha$ and $\beta$, subject to the constraint $\alpha^2+\beta^2=1$. Let us assume $\alpha$, $\beta\geq 0$. For fixed impurity functions, we are thus dealing with a family of models uniquely specified by the real constant $\alpha$. \textcolor{black}{In some theories with a Bogomol'nyi bound, the BPS property restricts the parameters of the model to a specific case (e.g., critical coupling for Maxwell-Higgs vortices or zero Higgs mass for Yang-Mills-Higgs monopoles). This does not occur here: as in the one-dimensional case~\cite{AdamI,AdamII}, infinitely many possibilities, corresponding to the range $\alpha\in[0,1]$, are possible}. These choices correspond to different couplings between defect and impurities. If one chooses  $\alpha=\beta=1/\sqrt{2}$, representing the case in which both terms in \eqref{TS} contribute equally to $E_s$, the first-order equations become
\begin{equation}\label{eq3}
\pb{1}\phi= W_{\phi}+\sigma_1,\;\;\;\;\;    \pb{2}\phi= W_{\phi}+\sigma_2.
\end{equation}
 A particularly simple class of models is given by the condition $\sigma_1=0$, wherein only one impurity is at play. This represents an asymmetry between directions $x$ and $y$, which is natural when inhomogeneities are present in the system. The potential may be taken as that of the simple and well-known $\phi^4$ model, with $W_{\phi}=1-\phi^2$. Eqs.~\eqref{eq3} form a system of  partial differential equations, which cannot in general be solved in closed-form. However, exact solutions can be found for some impurities. For example, the choice $\sigma_2=-2\sech^2(x-y)$ gives the defect $\phi=\tanh(x-y)$, depicted in Fig.~\ref{fig1} (left). This solution is a domain wall that behaves like a kink in the $x$ direction and like an antikink in the $y$ direction. In general, one may always generate an exact solution with $\sigma_1=0$ and a family of functions $\sigma_2$, which may be determined algebraically by solving the remaining equation.
 
 The energy density of this system is simply $\rho=0$, as the negative and positive contributions to the energy density cancel each other out. This result is consistent with the minimum value $E=0$ implied by insertion of the boundary conditions $\phi(x\to \pm\infty,y)=\pm1$, $\phi(x,y\to \pm\infty)=\mp1$. These boundary conditions mean that regions of the system that are distant from the impurity {\it choose} a vacuum value independently, so that the solution looks like a vacuum field far from the position of the inhomogeneity.
 
 Another interesting example appears with the impurities given by $\sigma_1=(2ax-1)\,\sech^2(ax^2+by^2)$ and $\sigma_2=(2by-1)\,\sech^2(ax^2+by^2)$, where the real parameters $a$ and $b$ can be used to describe distinct situations. Solving the first-order equations, we find $\phi=\tanh(a x^2+ b y^2)$. The energy density is now $\rho=2(ax+by)\,\sech^4(ax^2+by^2)$. In the case with $a=1$ and $b=1$, the solution, which can be seen in Fig.~\ref{fig1} (right) is a radially symmetric configuration located in the $xy$ plane. In the same way, one could use the BPS equations in the $D\geq3$ case to look for other field configurations. $D=2$ and $D=3$ radially symmetric solutions that are at least metastable have been investigated before in the homogeneous setting~\cite{stable, effective,curved, stableDwall}, where the symmetry of the defect is one of the assumptions made in the search for solutions capable of evading Derrick's argument. However, we stress that radial symmetry is not an independent assumption in the above example, but a consequence of the Bogomol'nyi equations and the form of the impurity. In this sense, energy minimization for the given boundary conditions and impurities \textit{implies} radial symmetry in the solution, which is a novel feature from our model. Eqs.~\eqref{eq3} may also be seen as a cylindrically symmetric reduction of the $D=3$ case. Indeed, if one sets $\sigma_3=0$, the BPS equations in the $z$-direction imply $\pb{z}\phi=0$.
 
In our model ${\cal L}={\cal L}_{st}+{\cal L}_{im}$, the expression \eqref{condition},  together with the first-order equations \eqref{eq3}, can be used to generalize the {\it virial} theorem from the standard case to
 \begin{equation}
E_v+E_{\sigma}^{(0)}+E_p=\frac{1}{2}E_{\sigma}^{(1)},
 \end{equation}
 where $E_v=\frac12\int d^2x(\sigma_1^2+\sigma_2^2)$, while $E_{\sigma}^{(0)}$ and $E_{\sigma}^{(1)}$ aggregate, respectively, the contributions originating from the coupling of impurities with $\phi$ and its first derivatives. Furthermore, the scaling argument that follows from the Derrick's theorem imposes that Eq. \eqref{TS} vanishes. This is different from other models in higher spatial dimensions, since here one gets a vanishing lower bound which cannot be connected to some topological invariant. However, we have explicitly shown the presence of nontrivial stable solutions.
 
 The above virial may be used to deepen our understanding of the role of impurities in the circumvention of Derrick's argument. It is only through the energy of the terms $E_{\sigma}^{(k)}$ that a nontrivial solution is supported, as this virial theorem is far less restrictive than the one found by Derrick in the absence of impurities. If a (localized) impurity is located far from the region of interest, then those terms may be neglected in the given region, and we are led, through Derrick's theorem, to a vacuum solution. Thus, the position of the defect is constrained to the region in which the densities $\rho_{\sigma}^{(k)}$ contribute significantly to the total energy. We  emphasize that this does not mean  that our solution will decay into a homogeneous configuration at sufficiently great distances to the impurity since the impurity is always included when the totality of space is considered. However, this reasoning shows that the solution must look like a vacuum configuration when measurements are made at a considerable distance from the source of the homogeneity. This is perfectly illustrated in our examples. As seen in Fig.~\ref{fig1}, both solutions differ appreciably from the trivial vacua at a finite neighborhood of the origin, which was chosen as the center point for the impurities in our examples. This reasoning justifies the boundary conditions of the problem, which must be chosen in a way that assures homogeneous vacuum solutions at great distances from the inhomogeinities of the system. Such distances may be precisely defined by the requirement that $E_{\sigma}^{(0)}\sim E_{\sigma}^{(1)}\sim 0$ at a given energy scale.

Since we are proposing a novel class of models, let us investigate the stability of the $D-$dimensional solutions around small fluctuations, by taking $\phi({\bf x},t) = \phi({\bf x}) + \eta({\bf x},t)$, where $\phi({\bf x})$ is the solution of Eq.~\eqref{eomstatic} and $\eta({\bf x},t)$ denotes the perturbations. By substituting this into Eq.~\eqref{eom} we get, considering the linear contributions in $\eta$,
\begin{equation}
\pu\Pu\eta + \left[U_{\phi\phi} + \alpha_k\sigma_k \left(\frac{U_{\phi\phi}}{\sqrt{2U}}-\frac{U_\phi^2}{(2U)^{3/2}}\right)\right]\eta = 0.
\end{equation}
We then take the fluctuations in the form $\eta({\bf x},t) = \sum_i \eta_i({\bf x})\cos(\omega_i t)$. By doing so, the above equation becomes
\begin{equation}\label{stabE}
-\pb{k}\pb{k}\eta_i + \left[U_{\phi\phi} + \alpha_k\sigma_k \left(\frac{U_{\phi\phi}}{\sqrt{2U}}-\frac{U_\phi^2}{(2U)^{3/2}}\right)\right]\!\eta_i = \omega_i^2\eta_i.
\end{equation}
This is a Schr\"odinger-like equation in $D$ spatial dimensions with stability potential given by
\begin{equation}
V_{stab} = U_{\phi\phi} + \alpha_k\sigma_k \left(\frac{U_{\phi\phi}}{\sqrt{2U}}-\frac{U_\phi^2}{(2U)^{3/2}}\right).
\end{equation}
By considering that the static solution is compatible with the first order framework, we use $U = W_\phi^2$ to get
\begin{equation}\label{stab}
V_{stab}=2\left(W_{\phi\phi}^2 + W_\phi W_{\phi\phi\phi} + \frac{\alpha_k\sigma_k}{ \sqrt{2}}W_{\phi\phi\phi}\right).
\end{equation}
The presence of the first order equation \eqref{bps} allows us to write the stability equation \eqref{stabE} in terms of the first-order operators
\begin{equation}
S_k\! = -\pb{k} + \alpha_k\sqrt{2}W_{\phi\phi}\quad\!\text{and}\!\quad S^\dagger_k\! = \pb{k} + \alpha_k\sqrt{2}W_{\phi\phi},
\end{equation}
in the form $S_k^\dagger S_k \eta_i = \omega_i^2\eta_i$. When $\sigma_k=0$ for all k, the only possible solution of the Bogomol'nyi equations is the  homogeneous vacuum, so the above analysis is consistent with the well known fact that those trivial solutions are stable.  In general, this factorization ensures that the solutions which arise from Eq.~\eqref{bps} are at least meta-stable, since negative eigenvalues are absent from the stability equation. There still remains the possibility of zero modes appearing, which means that, if such modes exist, the field could in principle change to another solution of~\eqref{bps} with the same energy and subject to the same boundary conditions.

In order to investigate the zero modes, one must consider a small perturbation $\phi\to\phi + \psi$ and linearize the first-order equations~\eqref{bps}. Through this procedure we find that zero modes would need to satisfy the system of equations $\partial_{k}\psi=\alpha_k\sqrt{2}W_{\phi\phi}|_{\phi_0}\psi,$ where $\phi_0$ is a given solution of~\eqref{bps}. This amounts to solve the equation $S_k^\dagger S_k\eta_0=0$ for the previously defined $S_k$, with the notation $\eta_0=\psi$.  It should be clear that this system is not solvable in general due to it being a system of up to $D$ differential equations with only one unknown function. For concreteness, let us first consider the solution $\phi(x,y)=\tanh(x-y)$, found in our first example. In this case, $W_{\phi}=1-\phi^2$, so zero modes must satisfy $\partial_{k}\psi=-2\tanh(x-y)\psi$. The $k=1$ equation gives $\psi=e^{A(y)}\sech^2(x-y)$, for some as of yet undetermined function $A(y)$. By differentiating $\psi$ with respect to $y$ and imposing that $\psi$ satisfies the $k=2$ equation, one is led to the condition $dA(y)/dy=-4\tanh(x-y)$, which clearly cannot be satisfied by any choice of $A(y)$. We must thus conclude that no zero modes exist. This means that the defect is stable and, in fact, completely specified by the boundary conditions at infinity.

It is not always easy to do this calculation explicitly since it may be impossible to write the integral of $W_{\phi\phi}|_{\phi_0}$ in terms of elementary functions. We may, however, use a known identity from calculus to prove the absence of zero modes for a wide class of models. Since equations with $\alpha_k=0$ are trivial to this discussion, let us consider only the equations with $\alpha_k$ strictly greater than zero. This assumption allows us to define $\mu^k\equiv \sqrt{2}\alpha_k x^k$, with use of which one may write
	\begin{equation}\label{zeromodes}
	\frac{\partial\psi}{\partial\mu^k}=W_{\phi\phi}|_{\phi_0}\psi,
	\end{equation}
	where $\psi$ is treated as a function of the rescaled coordinates $\mu^k$. Note that in both of our examples, $\mu^k=x^k$. Indeed, the equality of partial derivatives of a multivariate function implies that its dependence on the arguments is additive. In other words, solutions of~\eqref{zeromodes} must satisfy $\psi=\psi(\xi)$, where $\xi=\mu_1+...\mu_N$, where $N$ is the number of equations in~\eqref{zeromodes}. Now we may add the equations in~\eqref{zeromodes} and make use of the chain rule to find $\partial_{\xi}\psi(\xi)=NW_{\phi\phi}|_{\phi_0}\psi(\xi)$. If $\psi$ is zero everywhere, there are no zero modes. If not, then this equation cannot be satisfied in all space unless $W_{\phi\phi}|_{\phi_0}$, and thus $\phi_0$ itself, only depend on the coordinates through $\xi$, that is
	\begin{equation}\label{cond}
			\phi_0(\mu_1,...,\mu_N)=\phi_0(\mu_1+...+\mu_N)\equiv \phi_0(\xi).
	\end{equation}
	In particular, for both examples considered in this paper, the absence of zero modes for both solutions is ensured by the fact that they are not of the form $\phi=\phi(x+y)$. This is in stark contrast with the one dimensional case, in which solutions of the first order equations are generally defined up to a constant parameter, thus giving rise to a Moduli space~\cite{AdamII}. The presence of a Moduli space is useful for many applications, most notably in investigations of scattering where it allows for the adiabatic approximation~\cite{AdamII,Manton}. In the present case, if one wishes to find a Moduli space rather than solutions that are completely specified by saturation of the BPS bound, the impurity functions must allow for a solution of the form $\phi=\phi(\xi)$. Determining the constraints that such a property would impose on the impurity functions, as well as proving their existence and examining the symmetries involved are very interesting matters which warrant future investigations. For our current purposes, we note that the previous demonstration of metastability coupled with the absence of zero modes suffices to prove linear stability for a wide class of solutions which include our examples, meaning that (i) such solutions attain the minimum energy compatible with the boundary conditions (as implied by the Bogomol'nyi procedure developed earlier) and (ii) small perturbations cannot cause these solutions to be transformed into other field configurations with the same energy.

In this work, we have investigated real scalar field systems of standard dynamics coupled to impurities, which are used to break translation invariance in the theory, thus providing us the mathematical tools necessary to study field configurations without the assumption of spatial homogeneity. We have seen that Derrick's theorem does not generally hold in the presence of impurities. This allows us to explore simple scalar field systems in three or more spacetime dimensions, without the need of a non-standard potential and with the possibility of BPS solutions, which did not exist in the homogeneous version of this theory. We have also conducted a complete stability analysis and found, under appropriate assumptions, the absence of zero modes for the first-order equations.

Traditionally, and mainly due to Derrick's theorem, investigations of defects in more than one spatial dimension have centered in gauge theories and other systems with a higher degree of complexity when compared to real scalar fields. However, our discoveries seem to indicate that these fields become a much more powerful and versatile tool when impurities are present. For this reason, it is our hope that the present work may inspire other investigations on the subject. Being a natural extension of standard field theories, the models discussed here may find applications in any of the many contexts in which scalar fields play a role. See Refs.~\cite{Kivshar,Shnir, Chervon} for reviews including applications of scalar theories in areas which include optics, condensed matter physics and cosmology. 
Given the versatility of scalar fields, one may also search for solutions with some kind of symmetry, such as the spherically symmetric domain walls which have found applications in gravitation and cosmology, as explored, for example, in Refs.~\cite{curved,stableDwall, ColemanII, Stewart,SphericalI,SphericalIV}. Our procedure may also guide investigations in curved spacetime in a way similar to the case considered in Ref. \cite{effective}. Here, extensions including impurities seem adequate to deal with more realistic descriptions of the inhomogeneous universe, in this case benefiting from the solutions of Einstein's equations contained in Ref. \cite{BE}, with direct use in applications in astronomy, for instance. Moreover, we can couple non-minimally the real scalar field to the Maxwell field (see, e.g., \cite{BMM1,BMM2} and references therein) to search for localized solutions in higher spatial dimensions in the presence of the electromagnetic degrees of freedom. 

As a distinct example of a venue of extension, we find it important to note that the results developed here can be extended to theories containing several real scalar fields. Such theories are also forbidden by Derrick's theorem in the impurity-free scenario, and the extra degrees of freedom may be useful in applications, in particular, in the case of localized solutions engendering internal structure. The study concerning non-relativistic fields are also of interest, and may be considered to investigate  multicomponent condensates and multilayered materials.

Our investigation unveils an interesting procedure to circumvent Derrick's Theorem and represents a step forward in the direction of constructing localized scalar-field structures in arbitrary spatial dimensions in the presence of impurities. The radially symmetric solution found in this work may motivate study of collisions in the plane, as in the scattering of kinks in the real line introduced in Ref. \cite{Aprl}, with the appearance of the spectral wall phenomenon. It may also be considered in a way similar to the scattering of vortices by impurities in Bose-Einstein condensates \cite{BEs} and in the gauged Ginzburg-Landau model \cite{Ashcroft}. It engenders other distinct possibilities of applications of practical use and can be extended in different manners, under the addition of extra degrees of freedom, including the Einstein, Maxwell, and the Einstein-Maxwell possibilities. 

\acknowledgements{The work is supported by the Brazilian agencies Coordena\c{c}\~ao de Aperfei\c{c}oamento de Pessoal de N\'ivel Superior (CAPES), grant No 88887.485504/2020-00 (MAL), Conselho Nacional de Desenvolvimento Cient\'ifico e Tecnol\'ogico (CNPq), grants No. 303469/2019-6 (DB), 306151/2022-7 (MAM) and No. 401991/2022-9 (MAL), Federal University of Para\'\i ba (UFPB/PROPESQ/PRPG) project code PII13363-2020 and Paraiba State Research Foundation (FAPESQ-PB) grant No. 0015/2019.}



\end{document}